\documentclass[12pt]{article}
\usepackage{times}
\usepackage{epsfig}
\usepackage{color}
\newcommand{\nn}{\nonumber}
\newcommand{\be}{\begin{eqnarray}}
\newcommand{\ee}{\end{eqnarray}}
\newcommand{\beq}{\begin{equation}}
\newcommand{\eeq}{\end{equation}}
\def\bea{\begin{eqnarray}}
\def\eea{\end{eqnarray}}
\def\Eq#1{Eq.~(\ref{#1})}

\begin{document}


\renewcommand{\thefootnote}{\fnsymbol{footnote}}
\begin{flushright}
    IFIC/09-63 \\
     \today
\end{flushright}
\par \vspace{10mm}
  
\begin{center}

{\Large \bf Heavy colored resonances in $t\bar t+\mathrm{ jet}$ at the LHC}

\vspace{8mm}

{\bf Paola Ferrario~\footnote{E-mail: paola.ferrario@ific.uv.es}} and
{\bf Germ\'an Rodrigo~\footnote{E-mail: german.rodrigo@ific.uv.es}}

\vspace{5mm}
Instituto de F\'{\i}sica Corpuscular, 
UVEG - Consejo Superior de Investigaciones Cient\'ificas, \\
Apartado de Correos 22085, E-46071 Valencia, Spain. \\

\vspace{5mm}
\end{center}

\par \vspace{2mm}
\begin{abstract}
The LHC is the perfect environment for the study of new physics 
in the top quark sector. We study the possibility of detecting signals 
of heavy color-octet vector resonances, through the charge asymmetry, 
in $t\bar t+$jet events. Besides contributions with the $t\bar t$ pair 
in a color-singlet state, the asymmetry gets also contributions which 
are proportional to the color factor $f_{abc}^2$. 
This process is particularly interesting for extra-dimensional models, 
where the inclusive charge asymmetry generated by Kaluza-Klein 
excitations of the gluon vanishes at the tree level. 
We find that the statistical significance for the measurement
of such an asymmetry is sizable for different values of the 
coupling constants and already at low energies.
\end{abstract}

\setcounter{footnote}{1}
\renewcommand{\thefootnote}{\fnsymbol{footnote}}


\section{Introduction}

The physics of the top quark is one of the most promising research fields 
at hadronic colliders such as the Tevatron at Fermilab or the 
Large Hadron Collider (LHC) at CERN. Its huge mass compared with the 
other quarks, and of the same order as the Higgs boson vacuum 
expectation value, suggests that it can play an important role in 
the electroweak symmetry breaking. Since its discovery at Tevatron in 1995, 
it has been extensively studied and its properties have been measured 
with better and better precision. However, the optimal environment 
to perform top quark measurements is the LHC, due to its high energy 
reach ($14$ TeV center-of-mass energy at full activity). 
At the LHC a great amount of top-antitop quark pairs will be produced, 
thus allowing to develop analyses with high statistic.
There, physics at the TeV scale will be widely explored, 
carrying to a better determination of the Standard Model (SM) 
as well as, possibly, discovery of new physics. 

\begin{figure}[hbt]
\begin{center}
\includegraphics[width=9cm]{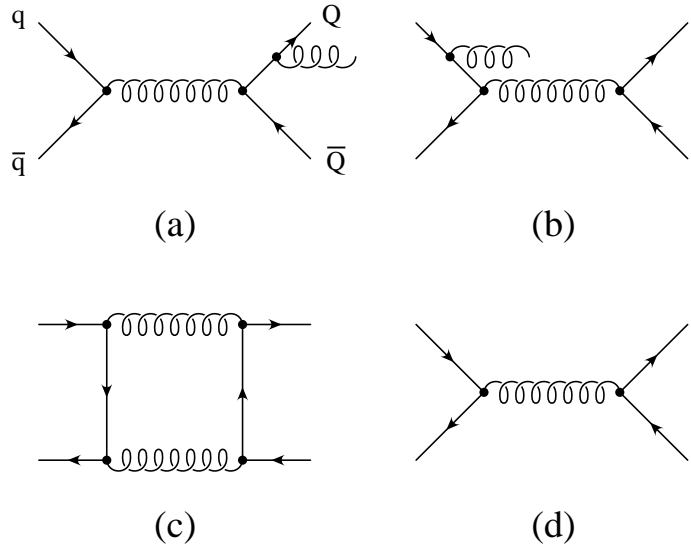}
\caption{Origin of the QCD charge asymmetry.\label{fig:aFBasym} }
\end{center}
\end{figure}

Several models predict the existence of heavy colored resonances 
decaying to top-antitop quark pairs, that in principle can be 
detected at the LHC, like axigluons~\cite{chiralcolor} and 
colorons~\cite{colorons} or Kaluza--Klein 
excitations in extra dimensional models~\cite{KK,Randall:1999ee,Dicus:2000hm,Agashe:2006hk,Lillie:2007ve,Djouadi:2007eg}. 
So far, masses under roughly $1$ TeV have been excluded 
by measurements performed at 
Tevatron~\cite{Aaltonen:2008dn,Aaltonen:2009qu,:2007dia,d0ttbar,graviton}. 
The natural signature of such resonances is to find a peak in the 
invariant mass distribution of the top-antitop quark pair.  
However, asymmetries can be an alternative way of revealing 
these resonances. In QCD, a charge asymmetry appears in the 
differential distributions for top quarks and antiquarks 
at $\mathcal{O}(\alpha^{3}_S)$ \cite{mynlo}. It is generated mostly 
by $q \bar q\longrightarrow t\bar t(g)$ processes, through the 
diagrams shown in Figure \ref{fig:aFBasym}. 
The contribution from $gq\longrightarrow t\bar t q$ is much smaller. 

The inclusive differential asymmetry at Tevatron is defined as: 
\be 
 A^{p\bar p}(\cos\theta) =\frac{N_t(\cos\theta)- N_{\bar{t}}(\cos\theta)}
{N_t(\cos\theta)+N_{\bar{t}}(\cos\theta)}\,,
\label{asymdef}
\ee 
and is equivalent to a forward-backward asymmetry. This leads to an 
integrated asymmetry of $4-5 \%$,with top quarks more abundant 
in the direction of the incoming 
proton~\cite{mynlo,Bowen:2005ap,Antunano:2007da,Almeida:2008ug}. 
The latest measurement of the forward--backward asymmetry 
by CDF attests \cite{newcdf}: 
\be 
A^{p\bar p} = 0.193 \pm 0.065_{\, \rm stat.} 
\pm 0.024_{\, \rm syst.}\,.
\label{eq:newcdf}
\ee
This $2\sigma$ discrepancy between data and theoretical prediction opens 
a window to the presence of new physics. Although it is too early to claim 
new physics there, because the statistical error of the measurement is 
rather large, this result clearly disfavours extra resonances with 
vanishing or negative contribution to the asymmetry, 
e.g. axigluons or colorons. 
As expected, this result has boosted a renovated 
interest in looking for new models that would account for this $2\sigma$ effect. 
In Ref.~\cite{Ferrario:2008wm} we have considered, in a model independent 
way, heavy color-octet boson resonances with arbitrary vector and 
axial-vector couplings to quarks, and pointed out that the 
charge asymmetry can be a better way of revealing new resonances than the 
total cross section, also at the LHC.
In Ref.~\cite{Ferrario:2009bz}, using the latest 
information from CDF on both the charge asymmetry and the invariant mass 
distribution~\cite{newcdf2}, we have set constraints on the couplings between 
quarks and colored resonances as a function of its mass. We found that in the 
flavor-universal scenario, a large value of the vector coupling is 
necessary in order to obtain a positive charge asymmetry at the Tevatron. 
A positive charge asymmetry can also be obtained in flavor-non-universal 
scenarios where the light and the top quarks couple to the heavy 
resonance with strengths of opposite sign.
A new model with this property has been 
constructed in~\cite{Frampton:2009ve}. 
Other flavor-universal axigluon-like models have been 
presented recently in~\cite{Carone:2008rx, Martynov:2009en,Zerwekh:2009vi}.  
The charge asymmetry in extra dimensional models has been 
analyzed in~\cite{Djouadi:2009nb}.
Different possibilities have also been explored in the $t$-channel:
diquarks~\cite{Arhrib:2009hu}, 
$Z$'/$W$' exchange with large flavor violating coupling~\cite{Jung:2009jz,Cheung:2009ch}, 
or scalar multiplets~\cite{Shu:2009xf}. 

At the LHC, due to its symmetric configuration, the integrated asymmetry 
vanishes. Nevertheless, it is still possible to find a charge asymmetry 
in suitable defined kinematic regions. Selecting events in a given 
range of rapidity in the central region, the integrated central charge 
asymmetry can be defined \cite{Antunano:2007da,Ferrario:2008wm}:
\beq
A_C(y_C) = \frac{N_t(|y|\le y_C)-N_{\bar{t}}(|y|\le y_C)}
{N_t(|y|\le y_C)+N_{\bar{t}}(|y|\le y_C)}~.\label{eq:acyc}
\label{eq:central}
\eeq
The central asymmetry $A_C(y_C)$ obviously vanishes if the 
whole rapidity spectrum is integrated, while a non-vanishing 
asymmetry can be obtained over a finite interval of rapidity. 

The production of top quark pairs together with one jet 
is important at the LHC: the exclusive cross-section for this process can 
reach roughly half of the total inclusive cross-section calculated 
at next-to-leading order (NLO)~\cite{Dittmaier:2008uj}. 
The asymmetry produced in $t\bar t+$jet by the interference 
of initial- with final-state real gluon emission 
(Figures \ref{fig:aFBasym}a and \ref{fig:aFBasym}b) is, obviously, 
a tree level effect, and moreover, one of the main contributions 
to the inclusive asymmetry.
In this paper we investigate the charge asymmetry in $t\bar t+$jet 
in the presence of a heavy color-octet vector resonance with different 
couplings to the quarks. We give the analytic form of the 
charge asymmetric contribution to the differential cross section, 
leaving the vector and axial-vector couplings as free parameters. 
We then concentrate on three different scenarios for such 
parameters and calculate the central charge asymmetry and its statistical 
significance at the LHC.

\section{Charge asymmetry at the LHC}

The LHC has already resumed its activity, after a one-year stop due 
to technical problems. It has started with an energy in the center-of-mass 
of a few TeV, in order to test the whole apparatus. 
In a second phase, the energy will rise to $7$ TeV and subsequently 
to $10$ TeV. Finally, the full $14$ TeV energy will be reached.
According to this planning, we have considered in our analysis 
both the center-of-mass energies of $7$ and $10$ TeV, 
in order to give predictions that can be tested in the first 
running period. 

\begin{figure}[tb]
\includegraphics[width=8cm]{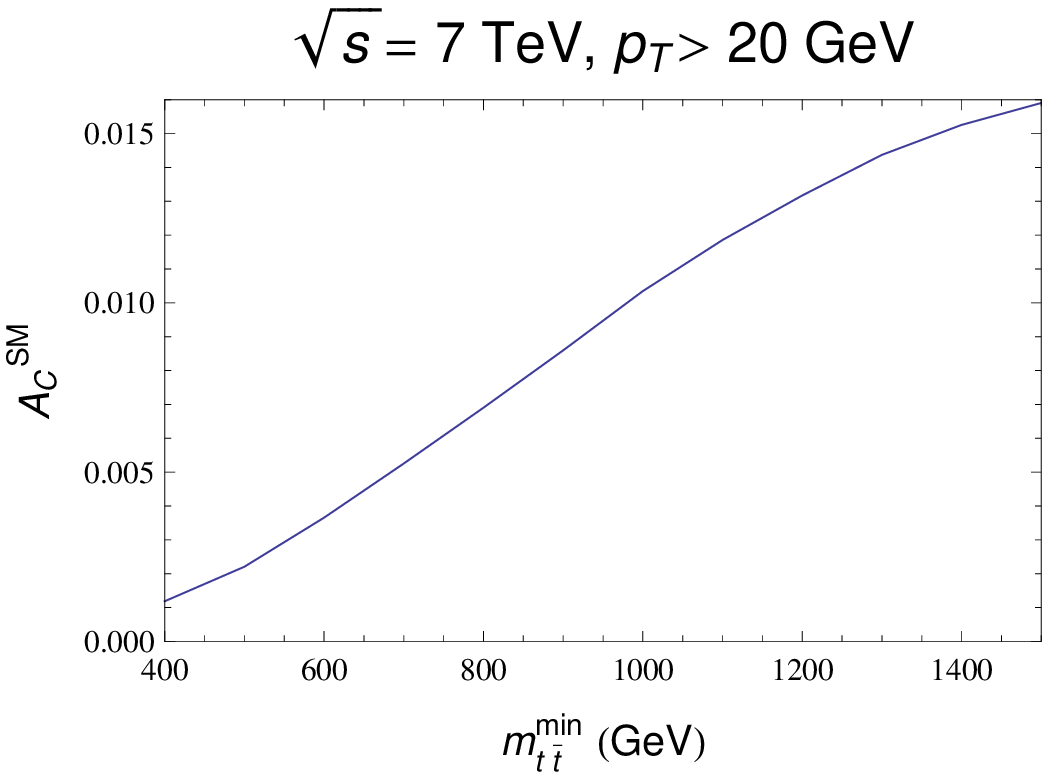}
\includegraphics[width=7.7cm]{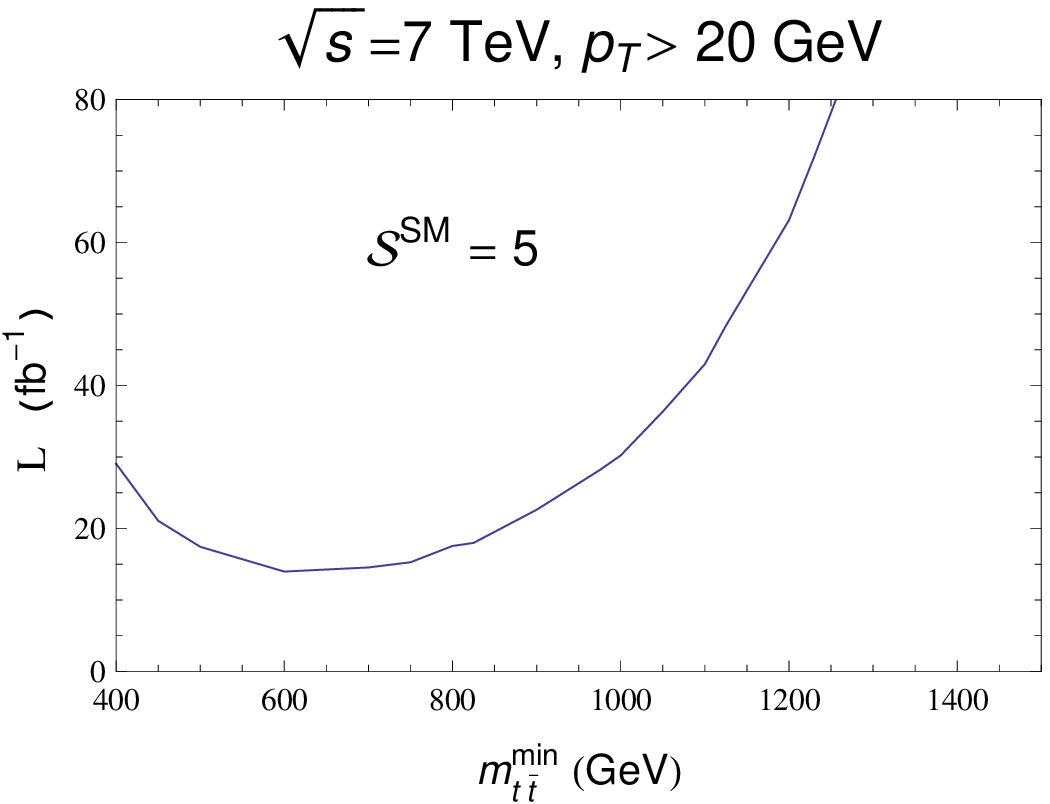}\\
\includegraphics[width=8cm]{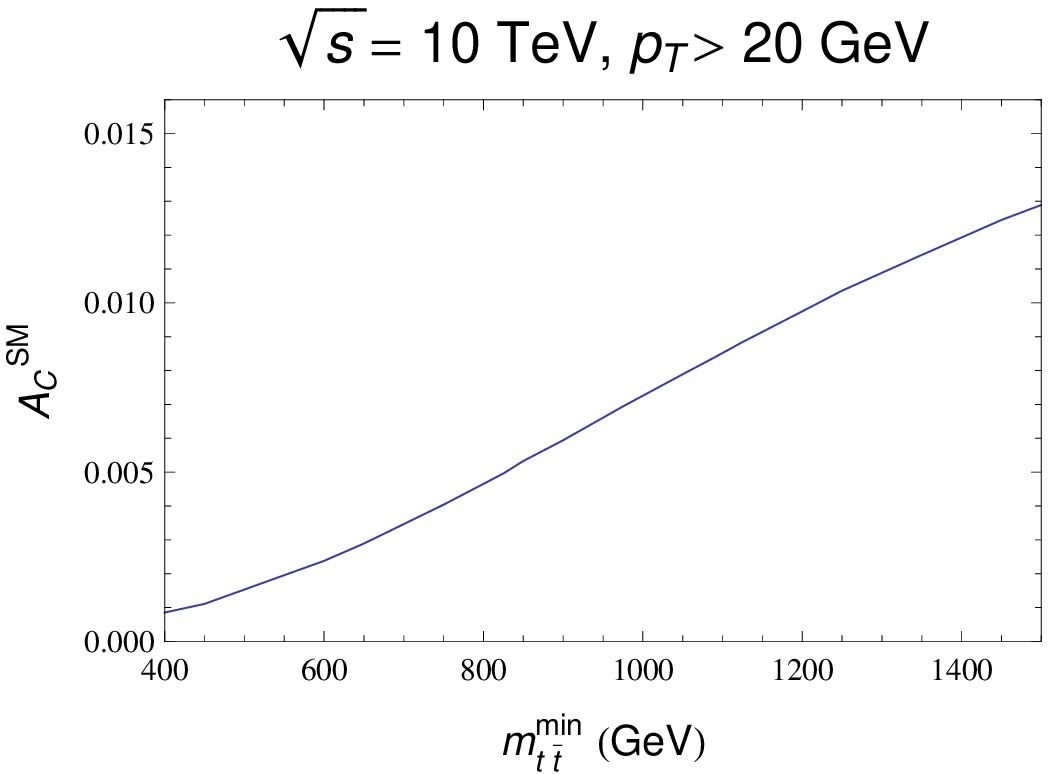}
\includegraphics[width=7.7cm]{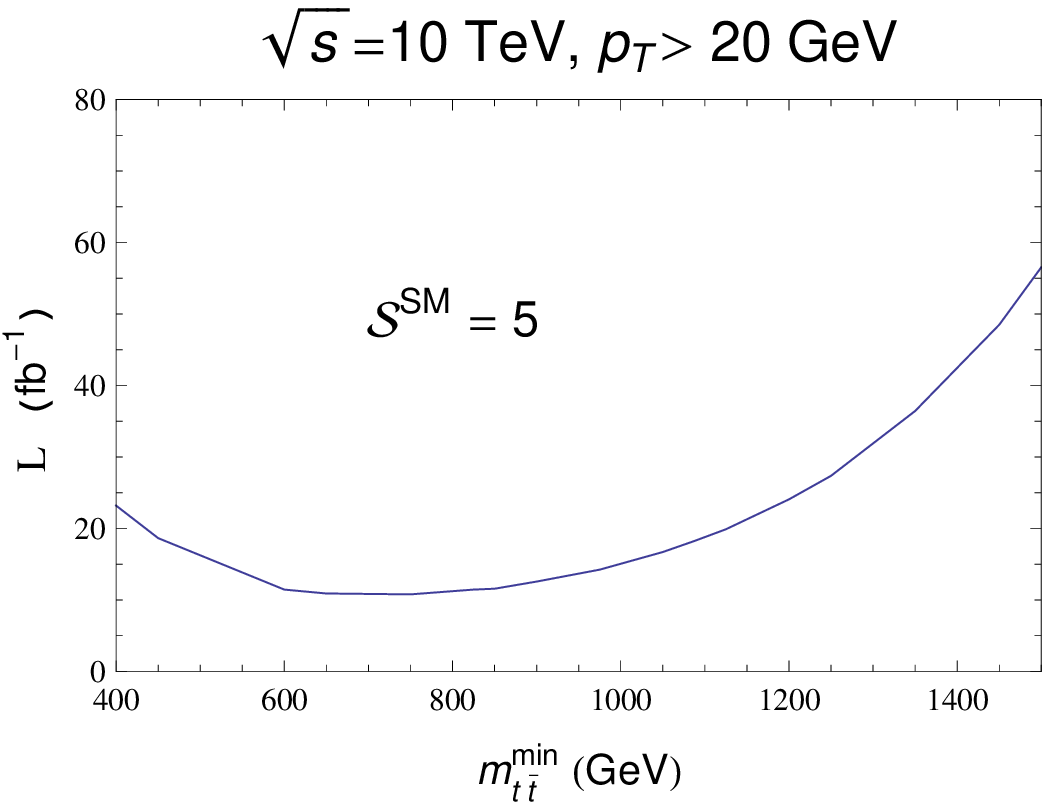}
\caption{Central charge asymmetry and luminosity at the LHC from QCD, 
as a function of the cut $m^{\mathrm{min}}_{t\bar t}$ for 
$\sqrt{s}=7$~TeV and $10$~TeV.\label{fig:qcd}}
\end{figure}

The SM predicts a charge asymmetry in $t\bar t+$jet already at tree 
level from $q\bar q$ events. This asymmetry is of similar size, 
but of opposite sign to the total $t\bar t$ inclusive asymmetry~\cite{mynlo}. 
At the LHC, however, top quark production is dominated by $gg$
fusion, which is charge symmetric. To reduce the contribution of 
these processes, and to enhance the asymmetry, it is necessary to 
perform a cut on the invariant mass of the top-antitop quark pair $m_{t\bar t}$.
In Ref.~\cite{Ferrario:2008wm} we found that for the central asymmetry
in \Eq{eq:central} values of the maximum rapidity around $y_C=0.7$ 
maximize the statistical significance. Thus, in the following, 
we fix $y_C=0.7$, and analyze the central asymmetry in the SM 
as a function of the cut on $m_{t\bar t}$. 
The additional jet is defined by using the $k_T$ 
algorithm~\cite{Catani:1992zp}, with minimum transverse momentum 
$p_T=20$ GeV and the jet parameter $R=0.5$ . 
In Figure \ref{fig:qcd} we show the results for center-of-mass  
energies of $7$ and $10$ TeV. We find that the asymmetry is positive 
and of the order of few percents (Fig.~\ref{fig:qcd}, left plots). 
As expected, at $7$ TeV the asymmetry is higher than at $10$ TeV, 
for the same value of $m_{t\bar t}^{\mathrm{min}}$, because the 
$q\bar q$ component is larger. The right plots in Fig.~\ref{fig:qcd} 
show the luminosity  that would be needed in order to have a statistical significance 
equal to $5$. The statistical significance ${\cal S^{\mathrm{SM}}}$  
of the measurement, defined as the number of standard deviations of 
which the asymmetry differs from zero, can be written in the following way:
\beq
{\cal S}^{\mathrm{SM}} = \frac{A_C^{\mathrm{SM}}}
{\sqrt{1-(A_C^{\mathrm{SM}})^2}} \, \sqrt{(\sigma_t+\sigma_{\bar t})^{\rm{SM}} 
\, {\cal L}}\simeq \frac{N_{t}-N_{\bar t}}{\sqrt{N_{t}+N_{\bar t}}}~, 
\eeq
where $\mathcal{L}$ is the total integrated luminosity. 
From Fig.~\ref{fig:qcd} we see that there is a minimum in the 
required luminosity for low values of $m_{t\bar t}^{\mathrm{min}}$. 
Before that minimum, $\mathcal{L}$ increases since the corresponding 
asymmetry approaches zero, while after the minimum, it increases 
because the number of events decreases. In conclusion, in the SM, 
with few tens of fb$^{-1}$ it would be possible to have a sizable 
significance for low values of $m_{t\bar t}^{\mathrm{min}}$. 
We should mention that we have not considered experimental 
efficiencies, therefore this number should be seen only as a 
lower limit. In a realistic analysis, much higher luminosities 
will be required to perform that measurement. However, we are 
interested here in showing the position of the minimum as 
a function of $m_{t\bar t}^{\mathrm{min}}$.

As in Ref. \cite{Ferrario:2009bz}, we consider now a toy model where 
a color-octet vector resonance can couple differently to light and 
top quarks. The vector and axial-vector couplings are denoted by 
$g_V^{q(t)}$ and $g_A^{q(t)}$, respectively, where the index $q$ 
indicates the light quarks and the index $t$ the top quarks. 
In Appendix \ref{ap:3jets} we list the expression for the 
asymmetric contribution to the $t\bar t+$jet differential cross section.
It is interesting to stress that, 
contrary to the SM, where top quarks contribute to the asymmetry  
only when they are in a color-singlet state (color factor equal 
to $d_{abc}^2$), we find also color-octet contributions proportional 
to the color factor $f_{abc}^2$.  We consider now three different 
scenarios. A large part of the parameter space for flavor-universal
couplings is disfavored because the inclusive asymmetry in that case 
is negative~\cite{Ferrario:2009bz}. In particular, axigluons such 
as originally introduced~\cite{chiralcolor}, 
i.\ e. with $g_V^{q(t)}=0$, $g_A^{q(t)}=1$, would be forbidden. 
Yet, it is possible to generate a positive inclusive asymmetry if 
the lighter quarks and the top quarks couple with different sign. 
Thus, as a first case, we examine a "modified axigluon", 
with $g_V^{q(t)}=0$ and $g_A^{t}=-g_A^{q}=1$. 
In the flavor-universal scenario, the only possibility that is 
still allowed at the $95\%$ C.L. is the one where $g_V$ takes 
high values and $g_A$ is constrained accordingly as a function 
of the resonance mass. So we choose as a second scenario
$g_V^{q(t)}=1.8$ and $g_A^{q(t)}=0.7$. 
In the third scenario we focus on a Kaluza--Klein gluon excitation 
in a basic Randall--Sundrum model: 
$g_V^q= -0.2~$,  $g_V^t= 2.5~$, $g_A^q=0~$, $g_A^t=1.5~$, 
as presented, for instance, in~\cite{Lillie:2007ve}. 
Since the axial coupling for the light quarks is zero, the 
inclusive central charge asymmetry vanishes at tree level. 
Thus, it is necessary to look at the hard 
emission process, where it becomes different from zero.
Accordingly, the inclusive charge asymmetry will get also non-vanishing 
loop contributions. 

The results for the asymmetry and the minimal luminosity to achieve 
a statistical significance of $5$ are shown in the 
Figures \ref{fig:7TeV} and \ref{fig:10TeV}. We have chosen $m_G=1.5$ TeV 
as a reference mass for the resonance. As in the pure QCD case, 
the maximal rapidity of $y_C=0.7$ is optimal to enhance the 
statistical significance, which is defined as:
\beq
{\cal S} = \frac{A_C-A_C^{\mathrm{SM}}}{\sqrt{1-(A_C^{\mathrm{SM}})^2}} \, 
\sqrt{(\sigma_t+\sigma_{\bar t})^{\rm{SM}} 
\, {\cal L}}~. \eeq
As expected, in the three models the asymmetry is slightly higher for 
$\sqrt{s}=7$~TeV. The luminosity required to have a fixed significance 
has a minimum for low values of $m_{t\bar t}^{\rm min}$, at around one half the mass 
of the resonance, for all the scenarios. In the flavor-universal 
case, we found that this minimum value is reached with even softer cuts. 
We find also that in this scenario the needed luminosity is lower than 
in the other two cases, and almost of about one order of magnitude less. A few hundreds 
of pb$^{-1}$ at relatively low values of $m_{t\bar t}^{\mathrm{min}}$ 
would allow a measurement in the first times of the LHC running. 
The Kaluza-Klein model shows an asymmetry of opposite sign compared 
to the other two cases. This can be an interesting way for distinguishing 
it from the other models. In Figs. \ref{fig:7TeV} and \ref{fig:10TeV} 
we also show the color-singlet contribution to the asymmetry. 
In the modified axigluon scenario, it has opposite sign compared 
with the total asymmetry. In the flavor-universal scenario it is about 
one half of the asymmetry. In the Kaluza-Klein model, the color-octet 
contribution is almost zero.

\section{Conclusions}

We have explored the central charge asymmetry in $t\bar t+\mathrm{jet}$ 
at the LHC. It receives contributions from top quark pairs both in a color-octet 
and in a color-singlet state. We have set a lower limit on the luminosity 
needed in order to have a statistical significance equal to $5$ for three 
different scenarios at $\sqrt{s}=7$ and $10$~TeV.
We have found that, in the flavor-universal case, 
this lower bound is around a few hundreds of pb$^{-1}$, 
while for the other scenarios few $\mathrm{fb}^{-1}$ are required.
These values depend, of course, on the resonance mass. 
For the three choices of the parameters that we have considered, 
the minimum of the required luminosity is reached for relatively
low values of $m_{t\bar t}^{\mathrm{min}}$. This is a non-trivial 
result as very boosted top quarks are difficult to distinguish 
from jets initiated by light quarks. 

NLO calculations of $t\bar t+$jet~\cite{ttjetnlo} in the SM 
show that the exclusive asymmetry is almost completely washed 
out at Tevatron. Although there is no reason why we should find 
the same behavior if a heavy resonance exists, it would be
interesting to extend this analysis at NLO, and to combine 
it with a realistic estimation of experimental efficiencies. 
From our analysis, the measurement of the charge asymmetry from $t\bar t+$jet 
events at the LHC seems promising, although challenging. Experimental analysis 
from the Tevatron with more statistics will also constrain further those 
resonances in the near future. 

\section*{Aknowledgements}

The work of P. F. is supported by the Consejo Superior de 
Investigaciones Cient\'ificas (CSIC). This work is also supported
by the Ministerio de Ciencia e Innovaci\'on under Grant
No. FPA2007-60323, by CPAN (Grant No. CSD2007-00042), 
by the Generalitat Valenciana under Grant No. PROMETEO/2008/069, 
and by the European Commission MRTN FLAVIAnet under Contract
No. MRTN-CT-2006-035482.

\appendix

\section{Asymmetric contribution to the $t\bar t+\mathrm{jet}$ cross section}
\label{ap:3jets}

The tree level cross section for $t\bar t$ production in the presence of a heavy 
resonance with arbitrary vector and axial-vector couplings to quarks, and the decay 
width of the resonance can be found in Ref.~\cite{Ferrario:2008wm}. 
We define the propagator of the heavy resonance as: 
\beq
G(s) = \frac{1}{s-m_G^2 + i \, m_G \, \Gamma_G}~.
\eeq

The charge asymmetric piece of the hard gluon radiation process 
\beq
q(p_1) + \bar{q}(p_2)\to Q(p_3) + \bar{Q}(p_4) + g(p_5)~,
\eeq
defined as: 
\beq
d\sigma_A^{q\bar q} \equiv \frac12
\left[ d\sigma(q\bar q \to Q X) - d\sigma(q\bar q \to \bar Q X) \right]~,
\eeq
is given by: 
\bea 
\frac{d\sigma_A^{q\bar q, hard}}{dy_{35}\, dy_{45} \, d\Omega} &=& 
\frac{\alpha_s^3 \, \hat s}{4\pi} \Bigg[ 
  \frac{d_{1}}{\hat s \, \hat s_{34}}   
+  \left( g_V^q \, g_V^t  \, d_1 - g_A^q \, g_A^t  \, f_{1} \right) 
{\rm Re} \left\{ \frac{G(\hat s_{34})}{\hat s} +
\frac{G(\hat s)}{\hat s_{34}}\right\} \nn\\ 
&-& 2 \, g_A^q \, g_A^t  \, f_{2} \, \frac{{\rm Re}\{G(\hat s)\}}{\hat s_{34}}
+ \left( g_V^q \, g_A^t \, f_{3} + g_A^q \, g_V^t \, d_{3} \right)
{\rm Im} \left\{  \frac{ G(\hat s_{34})}{\hat s}-
\frac{G(\hat s)}{\hat s_{34}}\right\}
\nn\\ 
&+& \left[ \left( (g_V^q)^2 + (g_A^q)^2 \right) 
\left((g_V^t)^2 \, d_1 + (g_A^t)^2 \, d_2 \right)
- 4 \, g_V^q \, g_A^q \, g_V^t \, g_A^t \, \left( f_{1} + f_{2} \right) \right]
{\rm Re}\{G(\hat s)^\dagger G(\hat s_{34})\} \nn\\ 
&+& 2 \, \left[ \left( (g_V^q)^2 + (g_A^q)^2 \right) \, g_V^t \, g_A^t \, f_{3}
+ g_V^q \, g_A^q \, \left( (g_V^t)^2 \, d_{3} + (g_A^t)^2 \, d_{4}\right) \right]
{\rm Im}\{G(\hat s)^\dagger G(\hat s_{34})\} \Bigg]
\nn\\ 
&-& (3\leftrightarrow 4)
\label{sigmaasym}
\eea
where 
\bea
&& 
{\rm Re}\{G(\hat s)\} = \frac{\hat s-m_G^2}{(\hat s-m_G^2)^2+m_G^2 \Gamma_G^2}~, \qquad 
{\rm Im}\{G(\hat s)\} = \frac{m_G \, \Gamma_G}{(\hat s-m_G^2)^2+m_G^2 \Gamma_G^2}~, \nn\\ &&
{\rm Re}\{G(\hat s)^\dagger G(\hat s_{34})\} = 
\frac{(\hat s-m_G^2)(\hat s-m_G^2)+m_G^2\Gamma_G^2}
{\left[(\hat s-m_G^2)^2+m_G^2 \Gamma_G^2\right]
\left[(\hat s_{34}-m_G^2)^2+m_G^2 \Gamma_G^2\right]}~, \nn\\  &&
{\rm Im}\{G(\hat s)^\dagger G(\hat s_{34})\} = 
\frac{(\hat s-\hat s_{34})\, m_G^2\, \Gamma_G^2}
{\left[(\hat s-m_G^2)^2+m_G^2 \Gamma_G^2\right]
\left[(\hat s_{34}-m_G^2)^2+m_G^2 \Gamma_G^2\right]}~,
\eea
and
\bea
d_{1} &=& \frac{c_1}{y_{35}} \left[
\left( \frac{y_{13}}{y_{15}}-\frac{y_{23}}{y_{25}} \right)
\left( y_{13}^2+y_{14}^2+y_{23}^2+y_{24}^2
+ 2m^2 \left(y_{34}+2m^2+y_{12}\right)\right) 
+ 4m^2 \left(y_{24}-y_{14}\right) \right]~, \nn\\ 
d_{2} &=& \frac{c_1}{y_{35}} \left[
\left( \frac{y_{13}}{y_{15}}-\frac{y_{23}}{y_{25}} \right)
\left( y_{13}^2+y_{14}^2+y_{23}^2+y_{24}^2
- 2m^2 \left(y_{34}+2m^2+y_{12}\right)\right) 
+ 4m^2 \left(y_{13}-y_{23}\right) \right]~, \nn \\ 
d_{3} &=& \frac{c_1}{y_{35}} \left[
\frac{y_{13}^2+y_{14}^2-y_{23}^2-y_{24}^2-2m^2 (y_{15}-y_{25})}{y_{15}\, y_{25}} \right]
 \frac{4}{\hat{s}^2} \, \epsilon^{p_1 \, p_2 \, p_3 \, p_4} ~,\nn\\ 
d_{4} &=& \frac{c_1}{y_{35}} \left[
\frac{y_{13}^2+y_{14}^2-y_{23}^2-y_{24}^2+2m^2 (y_{15}-y_{25})}{y_{15}\, y_{25}} \right]
 \frac{4}{\hat{s}^2} \, \epsilon^{p_1 \, p_2 \, p_3 \, p_4} ~, \nn\\  
f_{1} &=& \frac{c_2}{y_{35}} \Bigg[
\left( \frac{y_{23}}{y_{25}}-\frac{y_{13}}{y_{15}} \right)
\left( y_{13}^2+y_{14}^2+y_{23}^2+y_{24}^2 \right) \nn \\
&+& 4 \left( \frac{(y_{13}+y_{15}) y_{24} (y_{13}-y_{35})}{y_{15}} -
\frac{(y_{23}+y_{25}) y_{14} (y_{23}-y_{35})}{y_{25}}\right) \Bigg]~, \nn  \\
f_{2} &=& \frac{c_2}{y_{35}} \left[ 
2m^2 \left( y_{15}-y_{25} \right) \right]~, \nn \\
f_{3} &=& \frac{c_2}{y_{35}} \left[
\frac{y_{13}^2+y_{14}^2-y_{23}^2-y_{24}^2}{y_{15}\, y_{25}} \right]
 \frac{4}{\hat{s}^2} \, \epsilon^{p_1 \, p_2 \, p_3 \, p_4} ~,
\eea
with 
\beq 
y_{ij}=\frac{2p_i\cdot p_j}{\hat s}\, .
\eeq 
The colour factor are $c_1=\frac{d_{abc}^2}{16N_C^2}$, and $c_2=\frac{f_{abc}^2}{16N_C^2}$,
with $N_C=3$, $d_{abc}^2=2 C_F (N_C^2-4)=40/3$ and $f_{abc}^2=2C_F N_C^2=24$.

The charge asymmetric contribution of the flavor excitation process 
\beq
q(p_1) + g(p_2)\to Q(p_3) + \bar{Q}(p_4) + q(p_5)~,
\eeq
defined as: 
\beq
d\sigma_A^{q\bar q} \equiv \frac12
\left[ d\sigma(q g \to Q X) - d\sigma(q g \to \bar Q X) \right]~,
\eeq
is infrared finite and can be obtained just by crossing of 
the momenta $(2 \leftrightarrow 5)$ from \Eq{sigmaasym}.

\begin{figure}[!ht]
\begin{center}
\includegraphics[width=8cm]{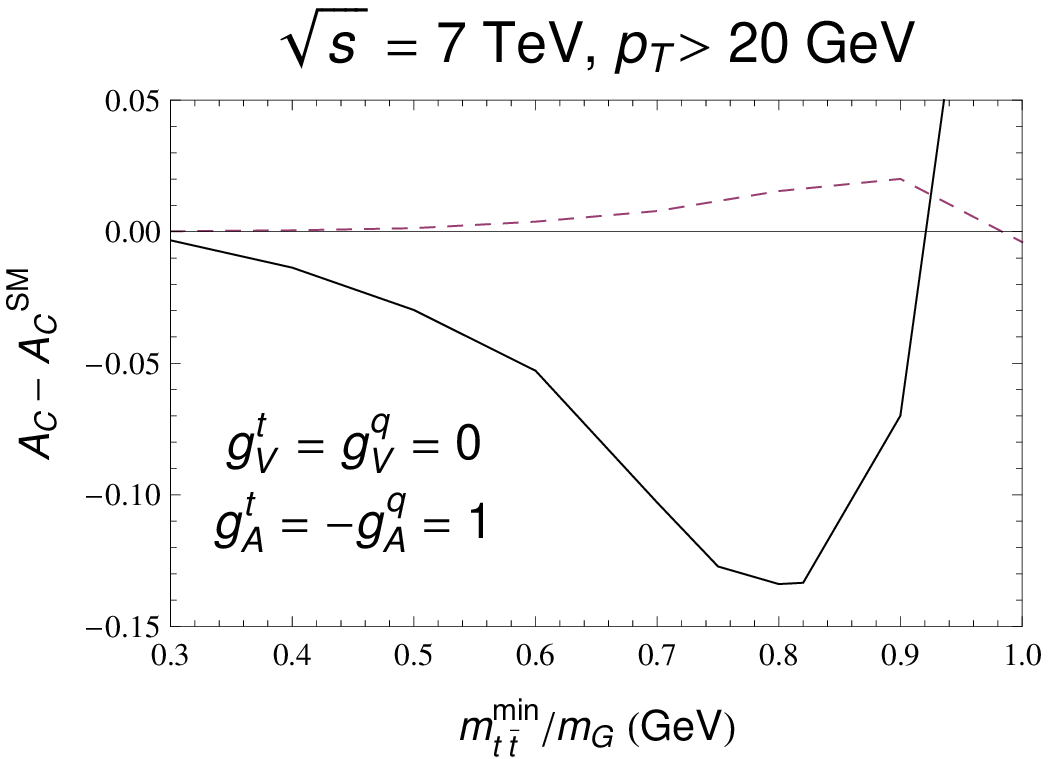}
\includegraphics[width=7.667cm]{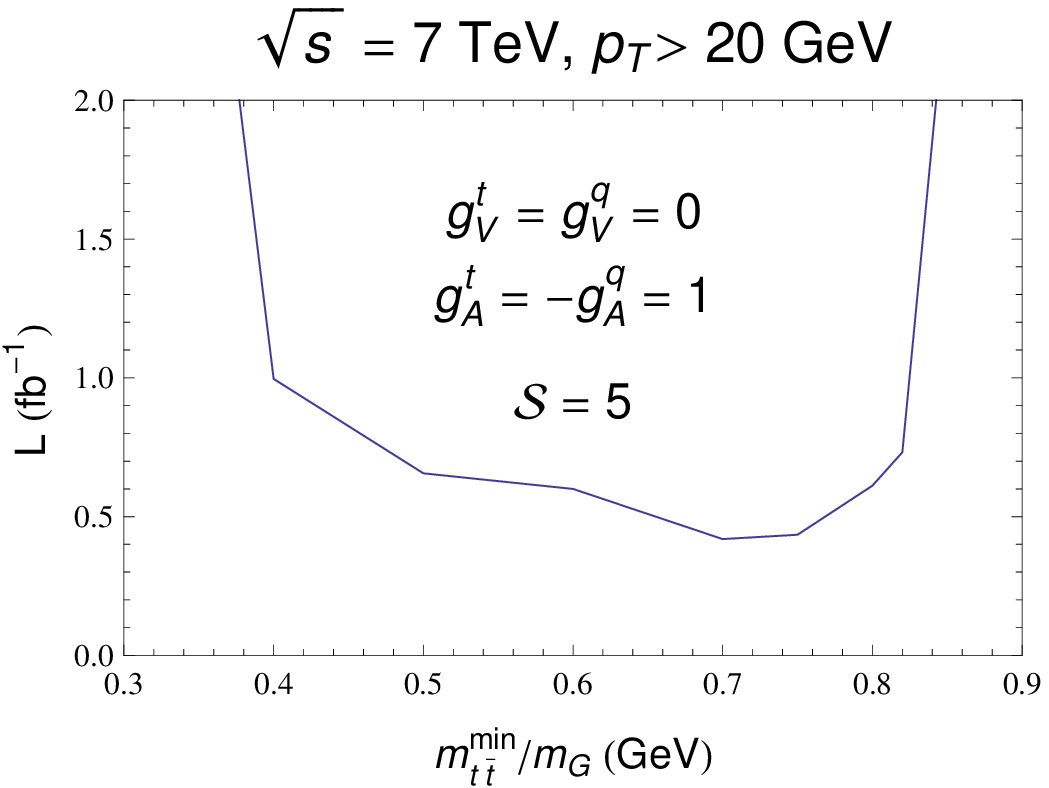}\\
\includegraphics[width=8cm]{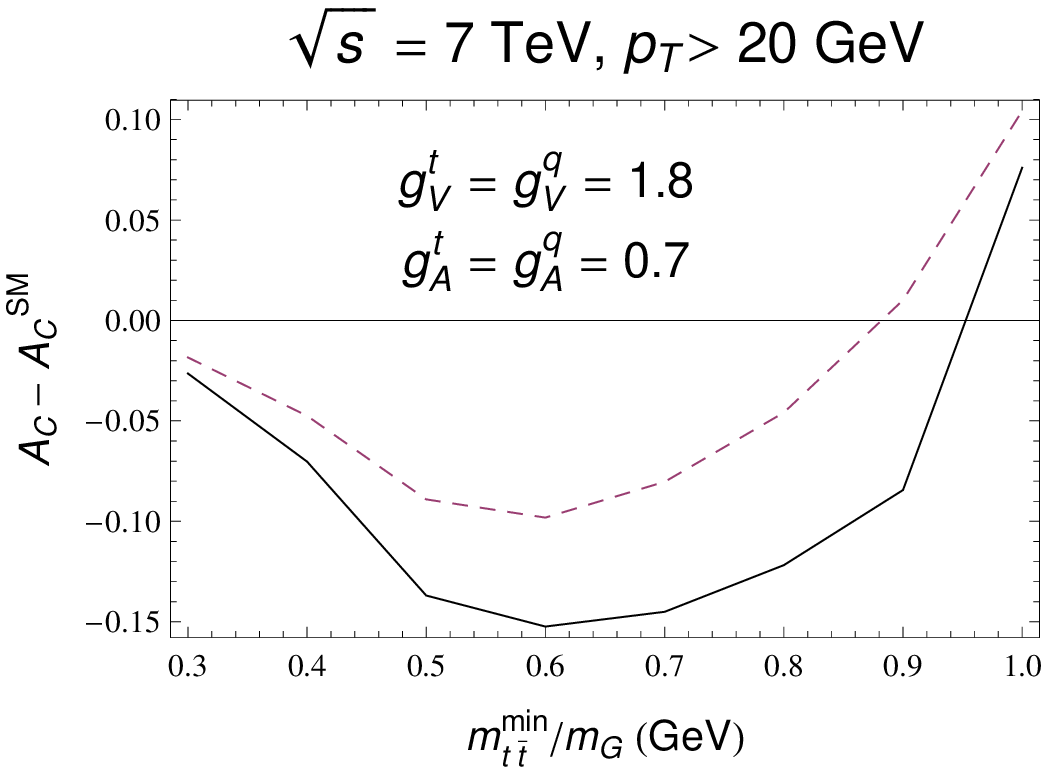}
\includegraphics[width=7.886cm]{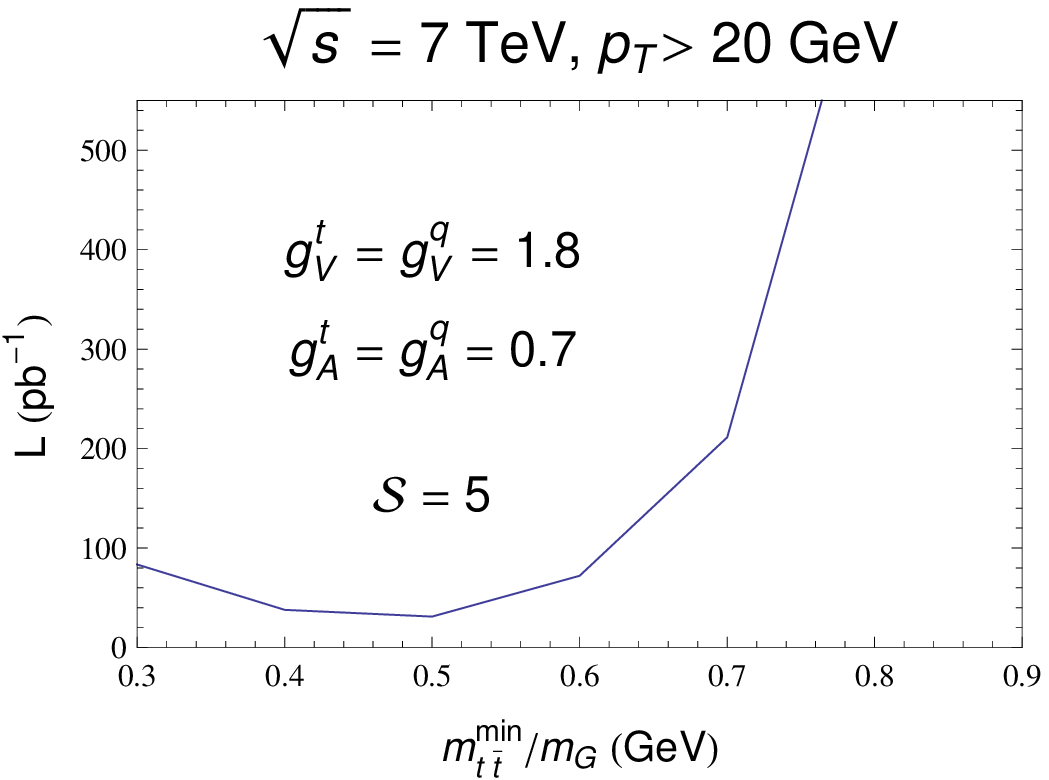}\\
\includegraphics[width=8cm]{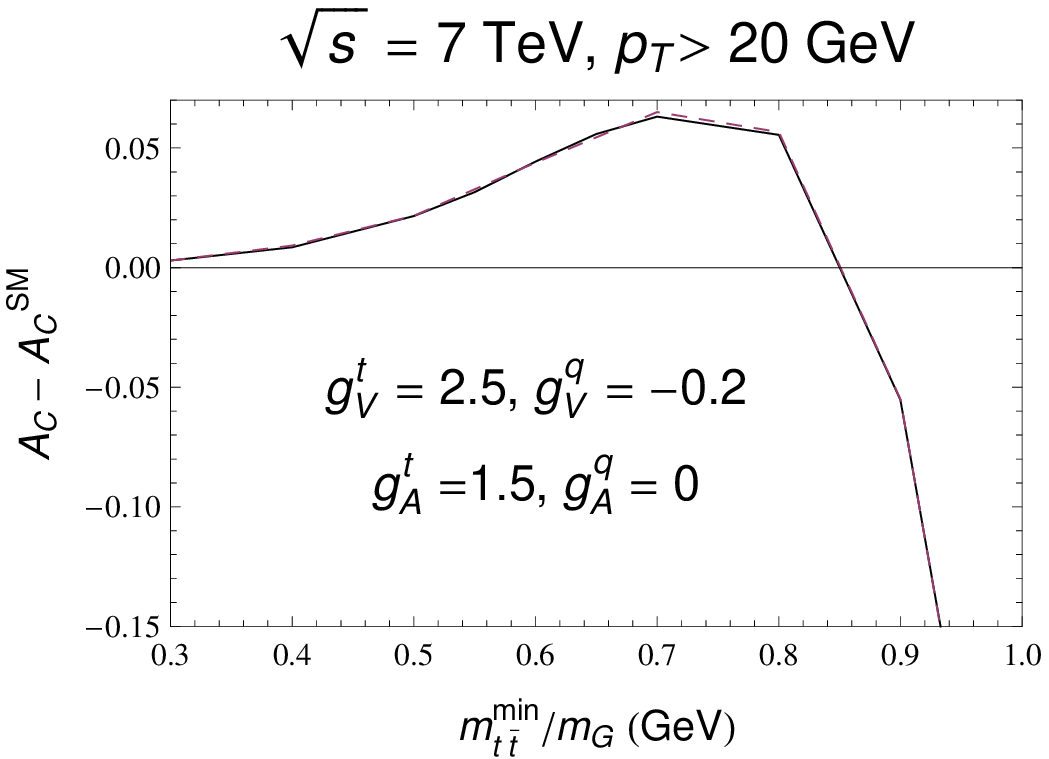}
\includegraphics[width=7.429cm]{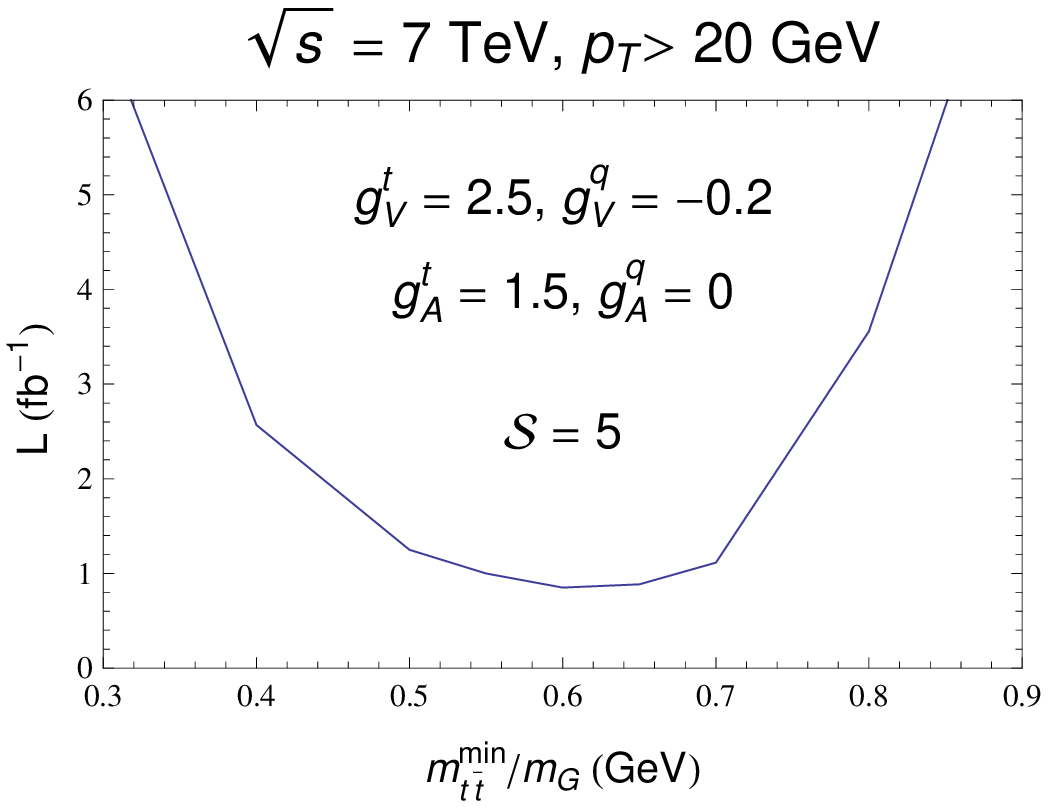}\\
\caption{Central charge asymmetry and luminosity to obtain a statistical 
significance $\mathcal{S}=5$ at the LHC, as a function of 
$m^{\mathrm{min}}_{t\bar t}$ for $\sqrt{s}=7$ TeV. 
The dashed line represent the contribution of the $d_{abc}^2$ terms. 
$m_G=1.5$ TeV. \label{fig:7TeV}} 
\end{center}
\end{figure}

\clearpage

\begin{figure}[!ht]
\begin{center}
\includegraphics[width=8cm]{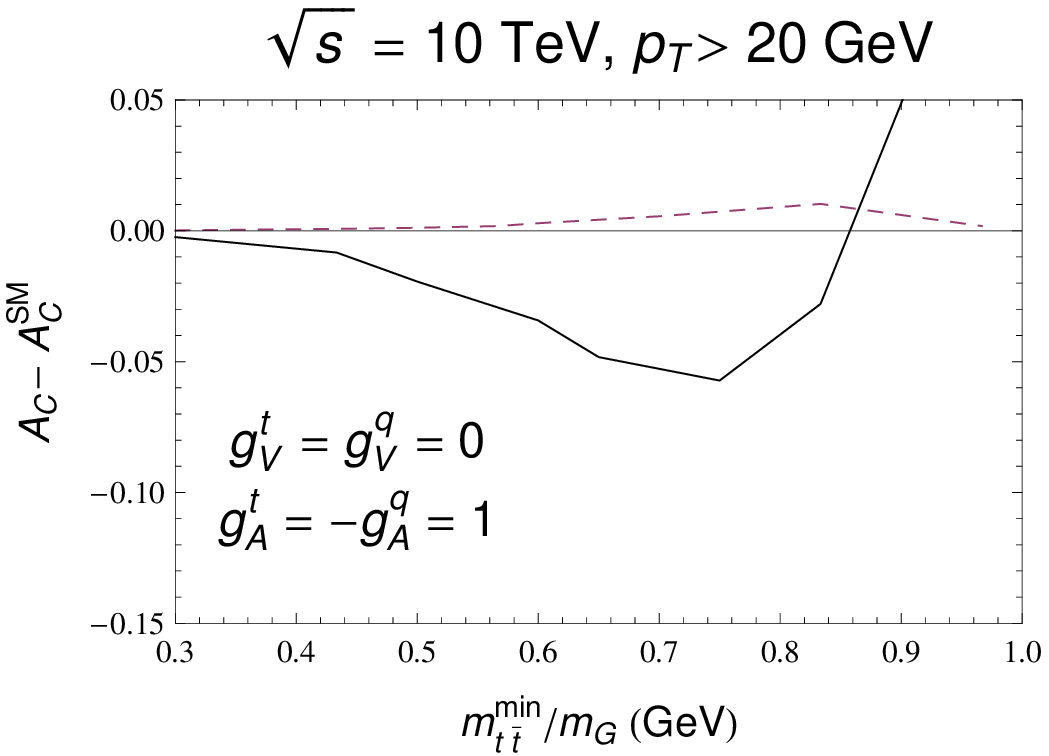}
\includegraphics[width=7.657cm]{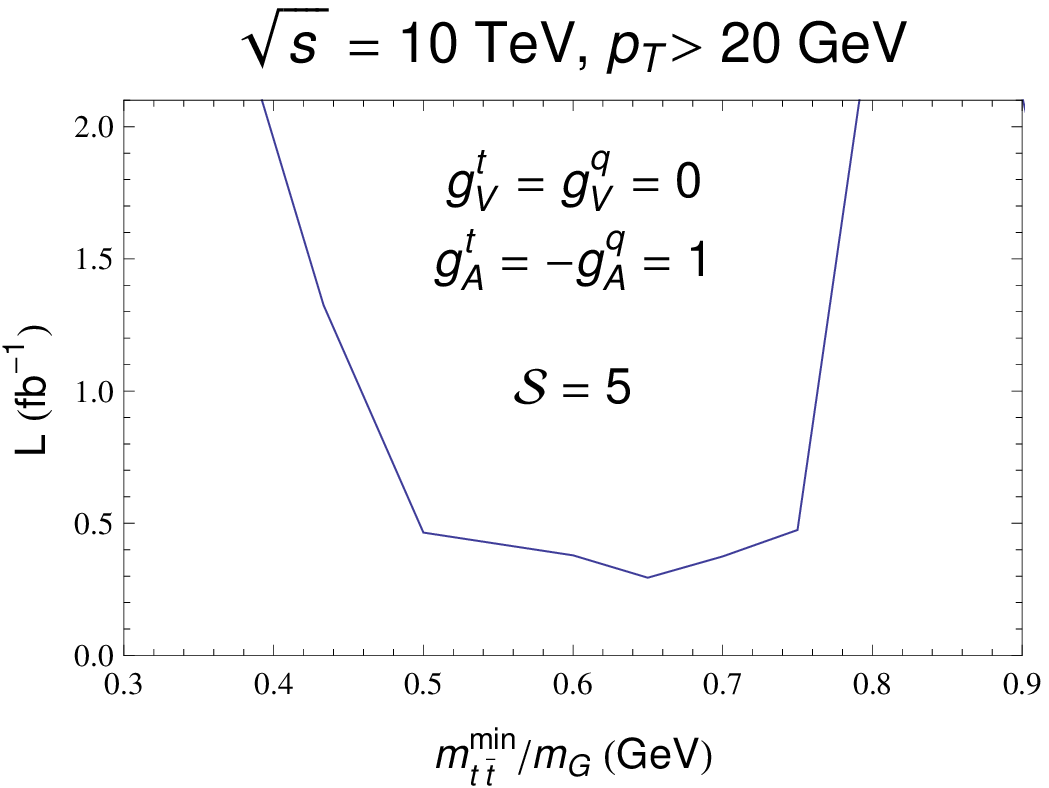}\\
\includegraphics[width=8cm]{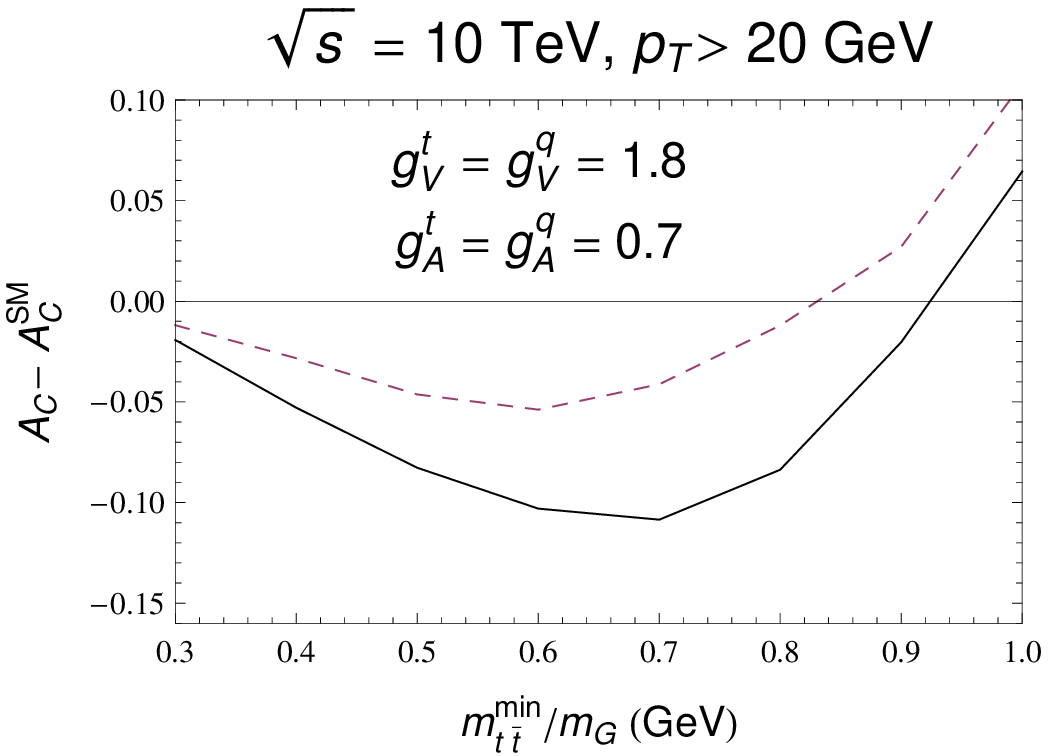}
\includegraphics[width=7.667cm]{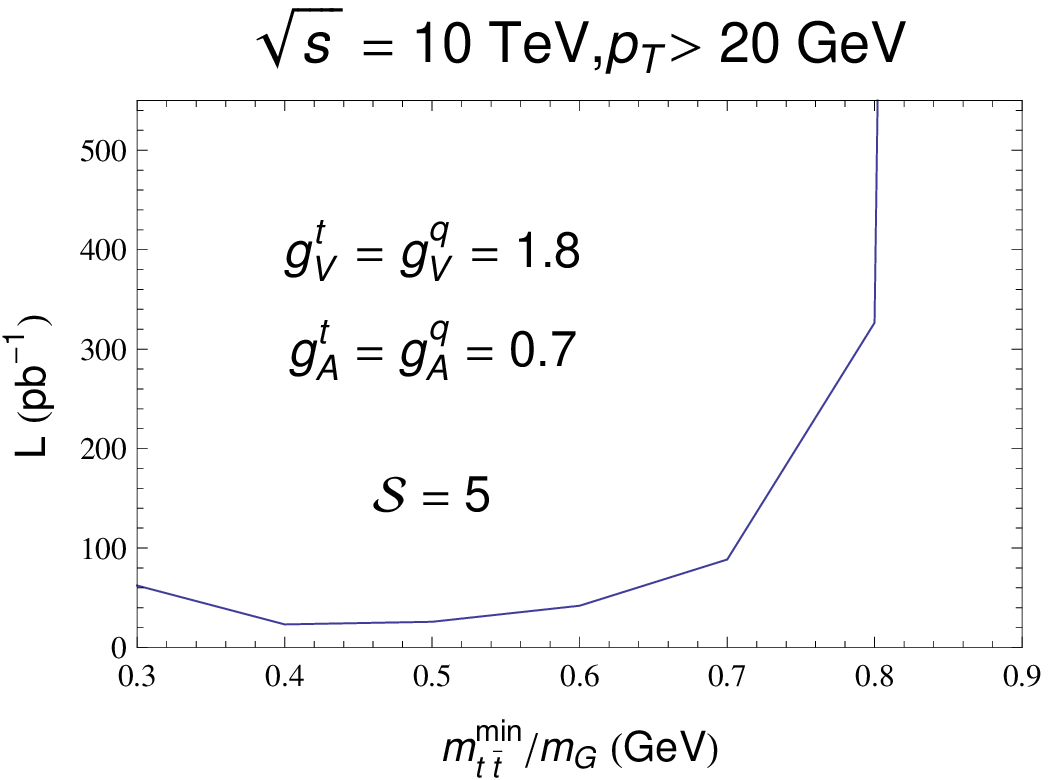}\\
\includegraphics[width=8cm]{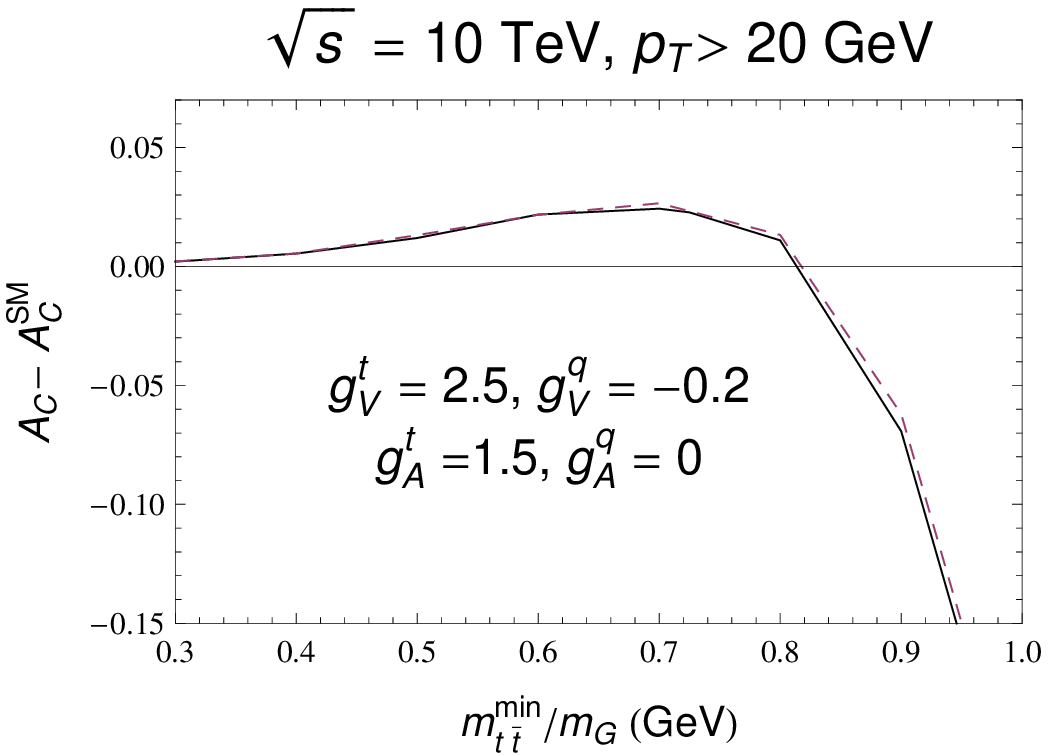}
\includegraphics[width=7.314cm]{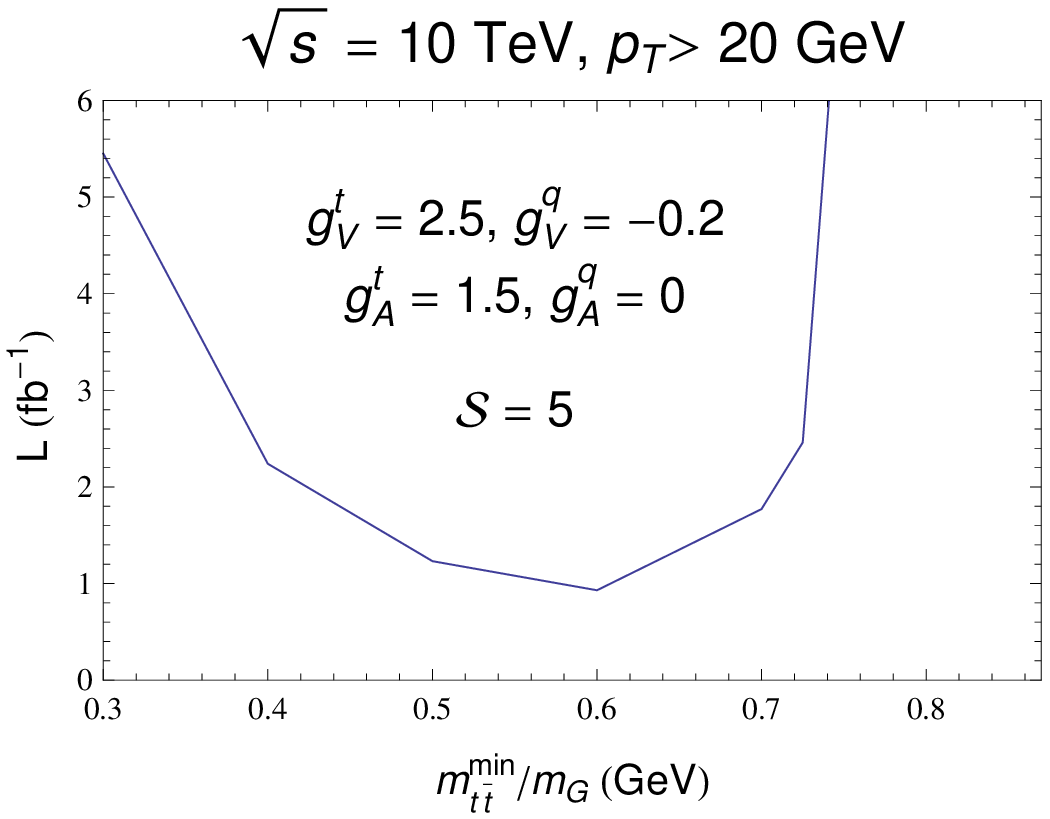}\\
\caption{Central charge asymmetry and luminosity to obtain a statistical 
significance $\mathcal{S}=5$ at the LHC, as a function of 
$m^{\mathrm{min}}_{t\bar t}$ for $\sqrt{s}=10$ TeV. 
The dashed line represent the contribution of the $d_{abc}^2$ terms. 
$m_G=1.5$ TeV.\label{fig:10TeV}}
\end{center}
\end{figure}


\clearpage


\begin{thebibliography}{90}



\bibitem{chiralcolor}
  J.~C.~Pati and A.~Salam,
  Phys.\ Lett.\  B {\bf 58} (1975) 333;
  L.~J.~Hall and A.~E.~Nelson,
  Phys.\ Lett.\  B {\bf 153} (1985) 430;
  P.~H.~Frampton and S.~L.~Glashow,
  Phys.\ Lett.\  B {\bf 190} (1987) 157;
  Phys.\ Rev.\ Lett.\  {\bf 58} (1987) 2168;
  J.~Bagger, C.~Schmidt and S.~King,
  Phys.\ Rev.\  D {\bf 37} (1988) 1188;
  F.~Cuypers,
  Z.\ Phys.\  C {\bf 48} (1990) 639. 

\bibitem{colorons}
  C.~T.~Hill,
  Phys.\ Lett.\  B {\bf 266} (1991) 419;
  C.~T.~Hill and S.~J.~Parke,
  Phys.\ Rev.\  D {\bf 49} (1994) 4454;
  R.~S.~Chivukula, A.~G.~Cohen and E.~H.~Simmons,
  Phys.\ Lett.\  B {\bf 380} (1996) 92.

  
\bibitem{KK}
  T.~Kaluza,
  Sitzungsber.\ Preuss.\ Akad.\ Wiss.\ Berlin (Math.\ Phys.\ ) {\bf 1921} (1921) 966;
  O.~Klein,
  Z.\ Phys.\  {\bf 37} (1926) 895
  [Surveys High Energ.\ Phys.\  {\bf 5} (1986) 241].

\bibitem{Randall:1999ee}
  L.~Randall and R.~Sundrum,
  Phys.\ Rev.\ Lett.\  {\bf 83} (1999) 3370.

\bibitem{Dicus:2000hm}
  D.~A.~Dicus, C.~D.~McMullen and S.~Nandi,
  Phys.\ Rev.\  D {\bf 65} (2002) 076007.

\bibitem{Agashe:2006hk}
  K.~Agashe, A.~Belyaev, T.~Krupovnickas, G.~Perez and J.~Virzi,
  Phys.\ Rev.\  D {\bf 77} (2008) 015003;
  K.~Agashe, A.~Falkowski, I.~Low and G.~Servant,
  JHEP {\bf 0804} (2008) 027.


\bibitem{Lillie:2007ve}
  B.~Lillie, L.~Randall and L.~T.~Wang,
  JHEP {\bf 0709} (2007) 074;
  B.~Lillie, J.~Shu and T.~M.~P.~Tait,
  Phys.\ Rev.\  D {\bf 76} (2007) 115016.

\bibitem{Djouadi:2007eg}
  A.~Djouadi, G.~Moreau and R.~K.~Singh,
  Nucl.\ Phys.\  B {\bf 797} (2008) 1.


\bibitem{Aaltonen:2008dn}
  T.~Aaltonen {\it et al.}  [CDF Collaboration],
  Phys.\ Rev.\  D {\bf 79}, 112002 (2009).

\bibitem{Aaltonen:2009qu}
  T.~Aaltonen {\it et al.}  [CDF Collaboration],
  Phys.\ Rev.\ Lett.\  {\bf 103}, 041801 (2009).

\bibitem{:2007dia}
  T.~Aaltonen {\it et al.}  [CDF Collaboration],
  Phys.\ Rev.\  D {\bf 77}, 051102 (2008);
  Y.~Oksuzian {\it et al.} [CDF Collaboration],
  Conf. Note 9844 (June 2009).

\bibitem{d0ttbar}
  D0 Collaboration,
  Conf. Note 5882 (March 2009).

\bibitem{graviton}
  C.~Wang {\it et al.} [CDF Collaboration],
  Public Note 9730 (April 2009);
  A.~Boveia {\it et al.} [CDF Collaboration],
  Public Note 9640 (January 2009).


  
\bibitem{mynlo}
  J.~H.~K\"uhn and G.~Rodrigo,
  Phys.\ Rev.\  D {\bf 59} (1999) 054017;
  Phys.\ Rev.\ Lett.\  {\bf 81} (1998) 49. 

\bibitem{Bowen:2005ap}
  M.~T.~Bowen, S.~D.~Ellis and D.~Rainwater,
  Phys.\ Rev.\  D {\bf 73} (2006) 014008.
 
\bibitem{Antunano:2007da}
  O.~Antu\~nano, J.~H.~K\"uhn and G.~Rodrigo,
  Phys.\ Rev.\  D {\bf 77} (2008) 014003;
  G.~Rodrigo,
  PoS {\bf RADCOR2007} (2008) 010.

\bibitem{Almeida:2008ug}
  L.~G.~Almeida, G.~Sterman and W.~Vogelsang,
  Phys.\ Rev.\  D {\bf 78} (2008) 014008.
  
 \bibitem{newcdf}
  G.~L.~Strycker {\it et al.} [CDF Collaboration],
  Conf. Note 9724 (March 2009).
  
\bibitem{Ferrario:2008wm}
  P.~Ferrario and G.~Rodrigo,
  Phys.\ Rev.\  D {\bf 78} (2008) 094018;
  J.\ Phys.\ Conf.\ Ser.\  {\bf 171}, 012091 (2009).

\bibitem{Ferrario:2009bz}
  P.~Ferrario and G.~Rodrigo,
  Phys.\ Rev.\  D {\bf 80} (2009) 051701.
  
\bibitem{newcdf2}
  T.~Aaltonen {\it et al.} [CDF Collaboration],
  Phys. Rev. Lett. 102 (2009) 222003.
  

\bibitem{Frampton:2009ve}
  P.~H.~Frampton,
  arXiv:0910.0307 [hep-ph];
  P.~H.~Frampton, J.~Shu and K.~Wang,
  arXiv:0911.2955 [hep-ph].


\bibitem{Carone:2008rx}
  C.~D.~Carone, J.~Erlich and M.~Sher,
  Phys.\ Rev.\  D {\bf 78} (2008) 015001.

\bibitem{Martynov:2009en}
  M.~V.~Martynov and A.~D.~Smirnov,
  Mod.\ Phys.\ Lett.\  A {\bf 24} (2009) 1897.

\bibitem{Zerwekh:2009vi}
  A.~R.~Zerwekh,
  arXiv:0908.3116 [hep-ph].

\bibitem{Djouadi:2009nb}
  A.~Djouadi, G.~Moreau, F.~Richard and R.~K.~Singh,
  arXiv:0906.0604 [hep-ph].


\bibitem{Arhrib:2009hu}
  A.~Arhrib, R.~Benbrik and C.~H.~Chen,
  arXiv:0911.4875 [hep-ph].


\bibitem{Jung:2009jz}
  S.~Jung, H.~Murayama, A.~Pierce and J.~D.~Wells,
  arXiv:0907.4112 [hep-ph].

\bibitem{Cheung:2009ch}
  K.~Cheung, W.~Y.~Keung and T.~C.~Yuan,
  arXiv:0908.2589 [hep-ph].

\bibitem{Shu:2009xf}
  J.~Shu, T.~M.~P.~Tait and K.~Wang,
  arXiv:0911.3237 [hep-ph].

  
\bibitem{Dittmaier:2008uj}
  S.~Dittmaier, P.~Uwer and S.~Weinzierl,
  Eur.\ Phys.\ J.\  C {\bf 59}, 625 (2009)
     
\bibitem{ttjetnlo}
  S.~Dittmaier, P.~Uwer and S.~Weinzierl,
  Phys.\ Rev.\ Lett.\  {\bf 98}, 262002 (2007).
  
\bibitem{Catani:1992zp}
  S.~Catani, Y.~L.~Dokshitzer and B.~R.~Webber,
  Phys.\ Lett.\  B {\bf 285}, 291 (1992).



\end{thebibliography}
\end{document}